# Environmental sensitivity of *n-i-n* and undoped single GaN nanowire photodetectors

F González-Posada[1,*], R Songmuang[2], M Den Hertog[2], and E Monroy[1]

[1] CEA-CNRS Group "Nanophysique et Semi-conducteurs", INAC-SP2M, CEA-Grenoble, 17 rue des Martyrs, 38054 Grenoble Cedex 9, France.

[2] Institut Néel-CNRS, BP 166, 25 rue des Martyrs, 38042 Grenoble Cedex 9, France.

[*] Presently at Applied Physics Department, Chalmers University of Technology, SE 412 96, Gothenburg, Sweden.

E-mail: fernando.gonzalez.posada@gmail.com

**Abstract**

In this work, we compare the photodetector performance of single defect-free undoped and *n-i-n* GaN nanowires (NWs). In vacuum, undoped NWs present a responsivity increment, nonlinearities and persistent photoconductivity effects (~ 100 s). Their unpinned Fermi level at the *m*-plane NW sidewalls enhances the surface states role in the photodetection dynamics. Air adsorbed oxygen accelerates the carrier dynamics at the price of reducing the photoresponse. In contrast, in *n-i-n* NWs, the Fermi level pinning at the contact regions limits the photoinduced sweep of the surface band bending, and hence reduces the environment sensitivity and prevents persistent effects even in vacuum.

Photodetectors based on semiconductor nanowires (NWs) picture new technological solutions for quantum electronics,[1] energy generation,[2] optoelectronics,[3] and sensorics.[4,5] The photodetector capabilities of single NWs are under study for a variety of materials,[6] including GaN.[7–10] For this application, III-nitride NWs present advantages in terms of physical and chemical robustness and bio-compatibility,[11] in addition to the photodetecting capabilities of a material family with a direct bandgap tunable in the ultraviolet (UV) / visible spectral range. We have previously reported that single $n$-$i$-$n$ NWs display photocurrent gain in the range of $10^5$–$10^8$, with the photocurrent scaling sublinearly with the excitation power.[10] Their spectral response, relatively flat for excitation above the GaN bandgap (< 360 nm wavelength), presents a visible rejection of more than six orders of magnitude, and their photocurrent time response is in the millisecond range. Although these results confirm that the photoresponse is dominated by the redistribution of charge at surface levels, these $n$-$i$-$n$ GaN NWs are insensitive to the measuring environment.

Given the large surface-to-volume ratio that characterizes the NW geometry, understanding the surface dynamics becomes a fundamental issue. From photoluminescence (PL) measurements in GaN-based NWs,[12,13] it is generally accepted that adsorbed oxygen reduces the nonradiative recombination time, resulting in a faster quenching of the NW emission in the air in comparison to experiments in vacuum, as also observed in GaN 2D layers[14] and ZnO NWs.[15] However, results become contradictory if we look at the photodetection properties: Some authors have stated that the photocurrent from GaN NWs is insensitive to the air/vacuum atmosphere,[10,16] whereas other groups report variations of the photocurrent in different gas environments.[8,17]

In this work, we compare the photodetector performance of single defect-free undoped and $n$-$i$-$n$ GaN NWs paying particular attention to the effect of the measuring environment (in vacuum or in the air). We conclude that the environment insensitivity of $n$-$i$-$n$ NWs with ohmic contacts is associated to the pinning of the Fermi level at the NW edges. In contrast, undoped NWs with Schottky contacts are significantly sensitive to the measurement atmosphere: In the air, adsorbed oxygen accelerates the surface processes, whereas in vacuum, due to the longer recombination time associated to the surface states, the behaviour of single NW photoconductors approaches the one of two-dimensional (2D) GaN photodetectors.

GaN NWs were grown by plasma-assisted molecular-beam epitaxy (PAMBE) on Si (111) under nitrogen-rich conditions, at a substrate temperature $T_S \sim 790$ ºC.[18] The NWs are defect free, they have a length of 1.2–1.5 µm and a diameter of 50−80 nm, and they exhibit N-polarity. The NWs under study are either undoped or *n-i-n* structures, the latter consisting of two Si-doped *n*-type edges and an undoped middle section with a nominal length of 400 nm. The residual doping in the PAMBE chamber is in the range of $\sim 1\times10^{17}$ cm$^{-3}$ for N-polar GaN. For *n*-type doping, the Si cell temperature was tuned to the value that yields a Hall electron concentration of $\sim 2 \times 10^{19}$ cm$^{-3}$ in 2D GaN layers.

Due to the large surface-to-volume ratio, the free carrier distribution in NWs is strongly affected by surface states. Applying the abrupt depletion approximation to a cylindrical structure,[19–21] total depletion of the NW occurs for a diameter below

$$d_c = \sqrt{\frac{16\varepsilon\varepsilon_0 \Phi}{e^2 N_d}}$$

where $\Phi$ is the surface potential (conduction band edge - Fermi level) and $N_d$ the doping density. Taking into account the diameter of the NWs under study, the typical values of surface potential in the literature (0.3 eV < $\Phi$ < 0.6 eV),[22–24] and the residual doping in our system, the undoped sections of the NWs should be fully depleted.

To perform the characterization of single objects, NWs were mechanically dispersed on heavily-doped Si substrates capped with a 280 nm thick SiO$_2$ layer. Single NWs were contacted using e-beam lithography, e-beam metallization and lift-off technique. The electrodes consisted of Ti/Al/Ti/Au (5/25/15/100 nm) for *n-i-n* NWs and Ni/Au (10/100 nm) for undoped NWs, in order to fabricate the ohmic and Schottky contacts, respectively. The right side of Figure 1 shows top-view scanning electron microscopy (SEM) images of *n-i-n* (top) and undoped (bottom) contacted NWs.

Current-voltage (*I*−*V*) characteristics were recorded using an Agilent 4155C semiconductor parameter analyzer. Photocurrent measurements were performed under bias ($V_b$) with the *n-i-n*

NWs connected to a load resistance ($R_L$ = 100 – 180 kΩ),[10] and the undoped NWs connected to a PDA200C current amplifier. Unless indicated, measurements were performed under chopped excitation using a TDS2022C oscilloscope or to an SR830 lock-in amplifier (input resistance = 10 MΩ). The spectral response was characterized using a 450 W Xe-arc lamp coupled to a Gemini 180 monochromator. At the exit, the light was directed with an optic fiber and focused onto the NW device. The variation of the photocurrent as a function of the optical power was analyzed using an Ar$^+$ laser (wavelength $\lambda$ = 244 nm, 488 nm) as excitation source. All the measurements presented in this work were performed at room temperature.

Figure 1 (left) compares typical *I–V* characteristics of single *n-i-n* and undoped NWs in the dark (measured in the air and in vacuum) and under UV illumination (in the air). Regarding the dark current, we observe a strong sensitivity to the atmosphere: Measurements in the air render current levels one order of magnitude lower than in vacuum, in agreement with previous reports[8] and with results in CdS NWs.[25] In the case of *n-i-n* NWs, the dark current displays the $I \sim V^{\alpha}$ ($\alpha \geq 2$) behavior characteristic of space-charge-limited transport, as expected due to the surface-induced depletion of the non-intentionally-doped NW sections. In the literature, space-charge limited current is observed for GaN NWs with diameters below 200 nm,[7,21] as also reported for CdS,[25] GaAs,[26] or ZnO NWs.[6] On the other hand, undoped NWs present dark current densities at least two orders of magnitude lower than *n-i-n* devices. Their approximately linear *I–V* characteristic is explained by the trap filling process characteristic of insulators prior to the application of the voltage.[27] Under illumination, undoped NWs display larger on/off current contrast in comparison to *n-i-n* NWs. However, the overall photocurrent level in undoped NWs is smaller than the one of *n-i-n* NWs by at least one order of magnitude.

Figure 2 presents on the left side the variation of the photocurrent as a function of the laser excitation power for an undoped NW measured in vacuum (pressure < 10$^{-4}$ mbar) and in the air. The right side of figure 2 describes the photodetector gain[28] as a function of the measurement frequency for *n-i-n* and undoped NWs in vacuum and in the air. In comparison to *n-i-n* devices, undoped NWs display lower photocurrent gain. In both cases, the photocurrent scales sublinearly with the optical power following an $I \sim P^{\beta}$ law with $\beta < 1$.[10] Statistically, the sublinear trend is less marked in the case of undoped NWs, with higher values of $\beta$ (0.4 − 0.6 in undoped NWs and 0.2 − 0.5 in *n-i-n* NWs).[29] Furthermore, undoped NWs present a high sensitivity to the measurement atmosphere: Whereas for *n-i-n* NWs photocurrent measurements in vacuum and in the air are in the same range (±5 %), for undoped NWs the photocurrent measured in vacuum can be more than one order of magnitude higher than in the air. Despite the enhancement of the photocurrent in vacuum, its sublinear behavior with the optical power remains unchanged, i.e. $\beta$ remains constant within an error bar ±5 %. A similar enhancement of the photoresponse in vacuum has been reported in ZnO NWs,[30] and in GaN NWs grown along the *m*-axis.[17]

Figure 3 compares the spectral response of an undoped NW photodetector in the air and in vacuum. In the air, the spectral response of *n-i-n* and undoped NWs is similar, approximately constant for wavelengths below 360 nm and displaying a visible rejection of more than three orders of magnitude, in agreement with the data presented of Bertness *et al.*,[8] although some previous reports show distinct photocurrent features below the GaN band gap.[31,32] To verify the visible blindness, the NWs were exposed to 1 W of a 488 nm $Ar^+$ laser without observing any photoresponse, whatever the NW or the measuring atmosphere. These results are in contradiction with a previous report which shows a degradation of the GaN NW UV/visible contrast under vacuum,[8] although the presence and role of persistent effects in that measurement were not indicated. In our case, the enhancement of the photoresponse of undoped NWs in vacuum is accompanied by a shape modification of the spectral response at short wavelengths (energies above the GaN bandgap) with the appearance of a second peak around 280 nm. This difference is understood by taking into account the high absorption coefficient at short wavelengths, which restricts the light penetration to a few nanometers, thus increasing the sensitivity to surface states and adsorbed impurities.

The variation of the gain with the chopping frequency displayed in Figure 2(right) points out the presence of slow phenomena. To clarify this issue, time response measurements were performed under 3 V bias exciting with a Xe arc lamp (λ = 350 nm). Figure 4 presents photocurrent decays recorded after ~2 min of continuous excitation. As previously reported, *n-i-n* NWs present nonexponential dynamics with an initial decay constant around 4-10 ms.[10] The decay becomes faster and increasingly nonexponential for higher excitation power, but no significant difference is observed as a function of the NW measuring environment, either in vacuum or in the air. In the case of undoped NWs, their behavior is similar to *n-i-n* devices when they are operated in the air. However, persistent photoconductivity effects are activated under vacuum in the range of tens-hundreds of seconds. These persistent effects approach the behavior of 2D GaN photoconductors.[33]

In summary, comparing undoped and *n-i-n* GaN NW photodetectors, in both devices we find a decrease of the dark current in the air. Under illumination, they display a similar spectral response with large UV/visible contrast. However, undoped NWs present a dark current three orders of magnitude lower than *n-i-n* structures, a gain of about one order of magnitude lower, and a strong dependence of the measurement environment: In vacuum, undoped NWs react with an enhancement of the photoresponse, accompanied by stronger nonlinearities, and persistent photoconductivity effects in the range of tens-hundreds of seconds.

To understand these results, we must keep in mind that the behavior of GaN NWs is dominated by the *m*-face sidewalls. Theoretical and experimental studies show that the Fermi level is unpinned in clean *m*-plane GaN surfaces, and their more stable atomic configuration is Ga-terminated surface [34,35]. Therefore, the GaN NW sidewalls show three Ga dangling bonds, from its inner core ($4s^2p^1$), that do not find the subsequent N layer to finish the $sp^3$ hybridization. In the air, adsorbed oxygen passivates the surface reducing the amount of charges available for conduction and modifying the radial potential profile in the NW. Furthermore, it is generally accepted that adsorbed oxygen results in a decrease of the carrier lifetime that can be measured by PL.[12,13]

Under illumination, we observe a variation of the NW conductivity $\Delta\sigma$. Keeping in mind that the conductivity is given by $\sigma = en\mu$, where $e$ is the electronic charge, $n$ is the carrier density, and $\mu$ is the carrier mobility, a change in conductivity, $\Delta\sigma = e(\mu\Delta n + n\Delta\mu)$, can occur either due to change in the carrier concentration, $\Delta n$, or to a change in the carrier mobility, $\Delta\mu$. The carrier concentration should scale linearly with the excitation, in contrast with the observed nonlinear behaviour of the photocurrent. Moreover, the carrier lifetime measured by PL in the ns range[12] is in contradiction with the photocurrent decay times described above (millisecond times for *n-i-n* NWs and significantly longer for undoped NWs). We can hence conclude that the photodetector response is dominated by $\Delta\mu$, which is affected by the surface band bending and by the scattering associated to surface states. When light is switched off, the photogenerated carriers recombine at a rate that is accelerated in presence of oxygen. However, the current recovery involves also a rearrangement of the surface charge, which is at the origin of the slow photocurrent components.

The different photoresponse of *n-i-n* NWs and undoped NWs is assigned to the different location and behavior of the Fermi level at the surface. Surface states play a major role on the undoped NW photocurrent dynamics due to the unpinned Fermi level. In the case of *n-i-n* NWs, the pinning of the Fermi level close to the conduction band, due to the *n*-regions, and the reduction of the *m*-plane surface potential, due to residual silicon on the sidewalls in the undoped regions, reduce the photoinduced sweep of the Fermi level, preventing persistent effects and reducing the environment sensitivity. Therefore, an important projection of the present work is that the doping profile of the NW is a critical parameter to determine not only its performance as a photodetector, but also its functionalization capabilities for chemical sensor applications.

In conclusion, single GaN NWs present high photocurrent gain, with the photocurrent scaling sublinearly with the excitation power, and with a spectral selectivity larger than six orders of magnitude. In the case of undoped NWs with Schottky contacts, the unpinned Fermi level at the NW sidewalls enhances the role of surface states in the photodetection process. In vacuum, the desorption of oxygen from the sidewalls slows down the recovery process, which results in higher responsivity associated to persistent photoconductivity effects. In contrast, in the case of *n-i-n* NWs, the pinning of the Fermi level at the contact regions limits the photoinduced sweep of the surface band bending, thus reducing the sensitivity to the environment and preventing persistent effects even in vacuum.

We thank J. Dussaud, Y. Curé, and Y. Genuist for their technical support. Partial financial support from ANR-2011-NANO-027 "UVLamp" and EU ERC-StG « TERAGAN » (#278428) is acknowledged. F. G-P thanks Prof. P. Lefebvre for sharing the ref. 13.

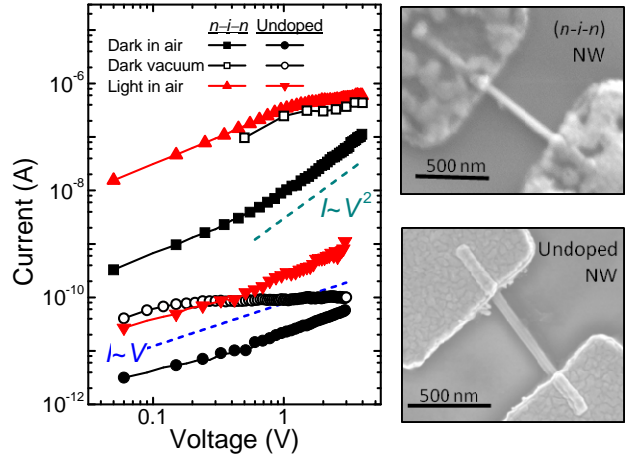

Figure 1. Left: $I-V$ characteristics of single *n-i-n* and undoped NWs measured in the dark (both in vacuum and in the air) and under UV illumination (in air). Black symbols correspond to the dark current; red symbols correspond to illumination with ~ 0.2 W/cm$^2$ of UV light ($\lambda$ = 244 nm). Right: Top-view SEM image of an *n-i-n* (top) and undoped (bottom) contacted single NW.

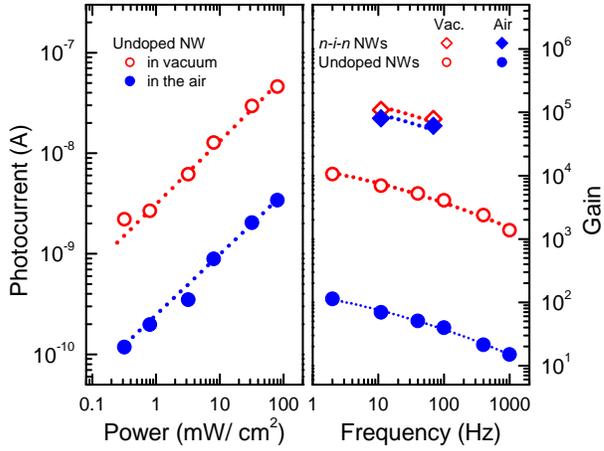

Figure 2. Left: Photocurrent variation as a function of the excitation power for a single undoped NW in vacuum and in the air. Measurements were made at 3 V bias and 11 Hz frequency. Right: Variation of the gain as a function of the measuring frequency of single *n-i-n* and undoped NWs under 3 V. The excitation power was 8 mW/cm$^2$. Dotted lines are guides for the eye.

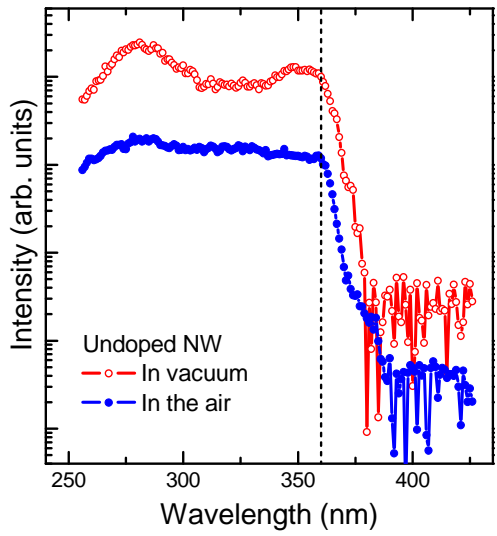

Figure 3. Spectral response of a single undoped NW in vacuum and in the air. Measurements were performed under 3 V bias, with a chopping frequency of 11 Hz, and scanning from long to short wavelengths.

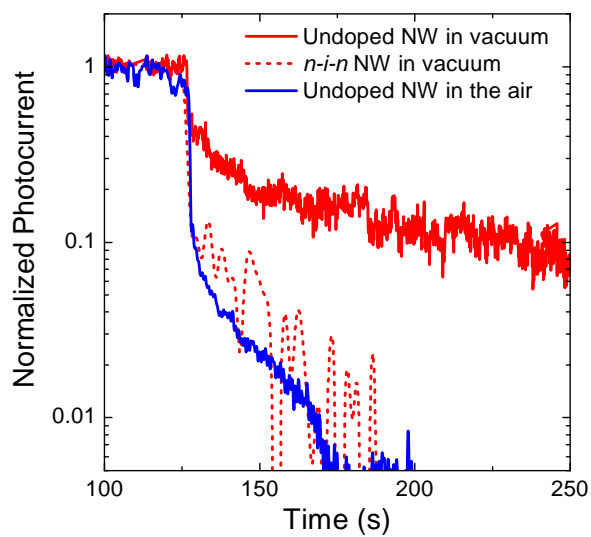

Figure 4. Photocurrent decays from single *n-i-n* and undoped NWs under 3 V bias after illumination with a Xe arc lamp (λ = 350 nm).

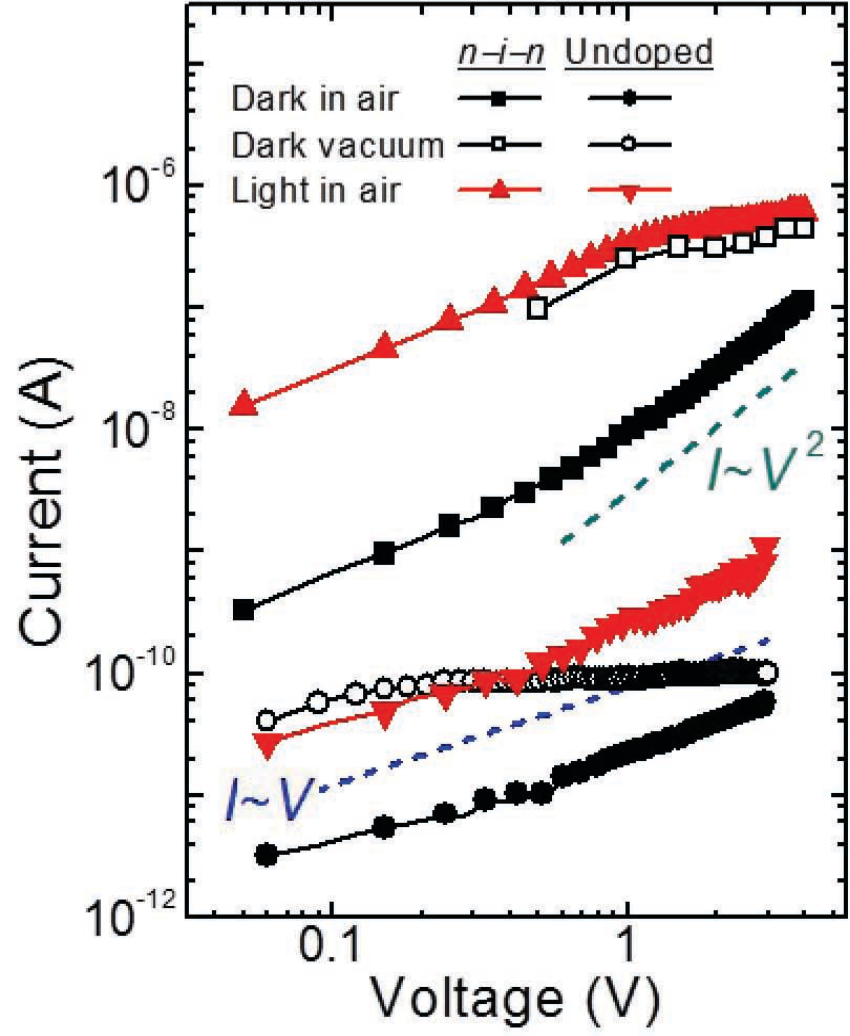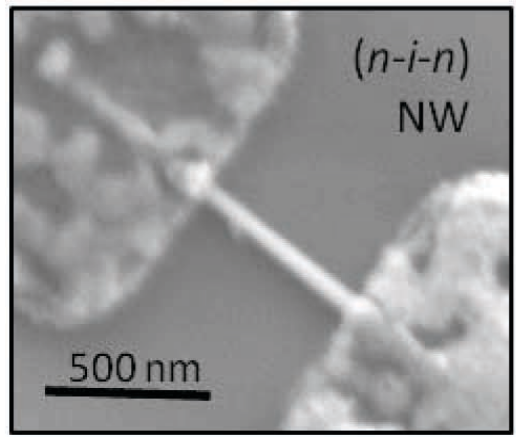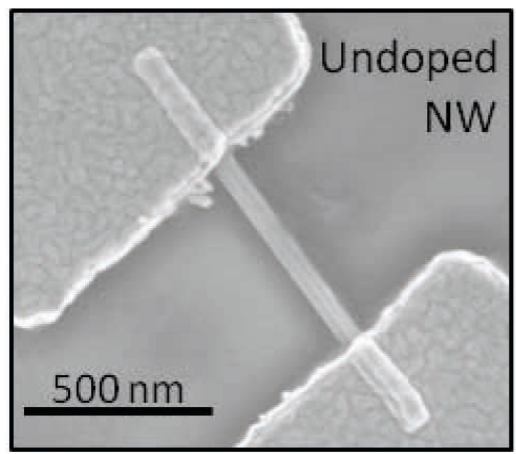

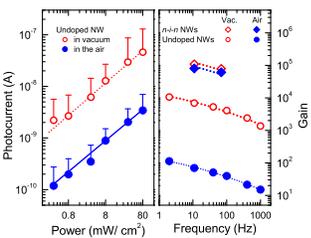

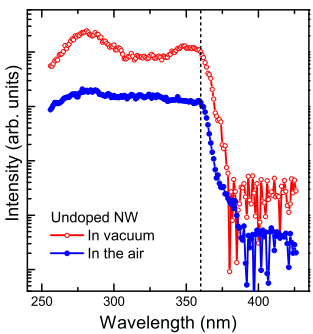

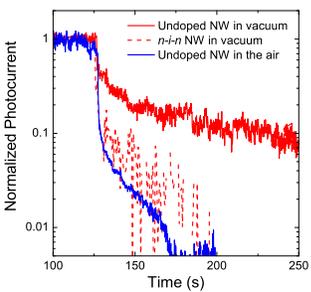